\documentclass[nofootinbib,superscriptaddress]{revtex4}
\usepackage{epsfig,amssymb,bm,graphics,color}
\usepackage{bbm}
\usepackage{slashed}              
\usepackage{amsmath}
\usepackage{xcolor}                
\usepackage{booktabs}

\DeclareMathOperator{\Tr}{Tr}

\begin{document}
\title{Regge Trajectories of 
Meson Excitations in the Pseudo-scalar and Vector Channels:
Exploring the Dyson-Schwinger -- Bethe-Salpeter Approach }
\author{R. Greifenhagen}
\affiliation{Institut f\"ur Theoretische Physik, TU Dresden, 01062 Dresden, Germany} \email{r.greifenhagen@hzdr.de}
\author{B. K\"ampfer}
\affiliation{Institut f\"ur Theoretische Physik, TU Dresden, 01062 Dresden, Germany}
\affiliation{Helmholtz-Zentrum Dresden-Rossendorf, PF 510119, 01314
Dresden, Germany} \email{kaempfer@hzdr.de}
\author{L. P. Kaptari}
\affiliation{Helmholtz-Zentrum Dresden-Rossendorf, PF 510119, 01314
Dresden, Germany}
\affiliation{Bogoliubov Lab.~Theor.~Phys., 141980, JINR, Dubna, Russia } \email{kaptari@theor.jinr.ru}

\begin{abstract}
The combined Dyson-Schwinger and Bethe-Salpeter equations
in rainbow-ladder approximation are used to search for Regge trajectories
of radially excited mesons in the pseudo-scalar and vector channels. We focus on the often employed
Alkofer-Watson-Weigel kernel which is known to deliver good
results for the ground state meson spectra;
it provides linear Regge trajectories in the $J^P= 0^-$ channel.

\end{abstract}

\maketitle

\section{Introduction}
\label{sec:1}
Despite of the apparent simplicity of the Lagrangian where Quantum Chromodynamics
(QCD) is based upon, it encodes an enormous richness of phenomena, most of them
related to the non-perturbative regime. While lattice QCD allows for an access to many
facets of the hadron spectra, the so called XYZ states pose still a challenge \cite{Yang:2016sws}.
Apart the quantitatively adequate description of low-lying
hadron states in various flavor channels, the higher excitations call also for a
description and confrontation with experimentally well established facts.
It is known for a long time that mesons of a given flavor composition can be grouped
on radial Regge trajectories according to
$M_n^2 = M_0^2 + n \mu^2$,
where $M_n$ stands for the mass (energy) labeled by
the radial quantum number $n = 0, 1, 2 , \cdots$,
$M_0$ denotes the ground state mass of a respective trajectory
and $\mu^2 = 1.25~{\rm GeV}^2$ \cite{Anisovich:2000kxa} or $1.35~{\rm GeV}^2$ \cite{Masjuan:2012gc}
is a universal slope parameter
(cf.\ \cite{Masjuan:2012gc,Masjuan:2013xta} for a recent account and
\cite{Klempt:2007cp,Brodsky:2014yha,Bugg:2012yt,Bugg:2004xu} for the discussion of the experimental data base).
More generally, Ref.~\cite{Afonin:2006wt} advocates an ordering according to
$M_{n, J}^2 = \hat a (n + J) + \hat c$, where $J$ stands for the angular momentum
and $\hat a$ and $\hat c$ are appropriate
constants, see also \cite{Afonin:2016wie}.
Often, a grouping according to $M_J^2 =M^2(0) + \hat \beta J$ is
considered prototypically for a linear orbital Regge trajectory.

While being a phenomenological ordering scheme, the arrangement of hadron states on Regge trajectories
should emerge from QCD, ideally directly without approximations or based on certain symmetries or as result
of suitable models. In fact, the relativistic quark model \cite{Ebert:2009ua,Ebert:2009ub} delivers such linear
trajectories. Also holographic models based on the AdS/CFT correspondence
(cf.~\cite{Brodsky:2014yha,Afonin:2009xi}) cope with Regge trajectories
\cite{Afonin:2009pd}, or even use them
as input for constraining the dilaton dynamics for further investigations \cite{Li:2013oda,Li:2012ay,Zollner:2017ggh}.
Moreover, functional formulations of QCD such as combined Dyson-Schwinger (DS)
and Bethe-Salpeter (BS) equations
address the issue of recovering Regge trajectories \cite{Fischer:2014xha,Hilger:2017jti}
with appropriate interactions kernels and truncation schemes \cite{Binosi:2016rxz}.
The latter approach is interesting since it provides the avenue towards addressing the important quest
for medium modifications of hadrons in a hot and dense hadron medium \cite{Wang:2013wk}.
Considering the medium created transiently in the course of relativistic heavy-ion collisions, the interplay
of confinement and chiral symmetry restoration poses further challenges \cite{Suganuma:2017syi}.
In this context, radial excitations of quarkonia play an important role
as diagnostic tool: The relative strengths of $\psi (2s)$ or $\Upsilon (2s)$
states to $J/\psi$ or $\Upsilon (1s)$, measured via their $e^+ e^-$ and
$\mu^+ \mu^-$ decay channels, depend on the centrality in nucleus-nucleus
collisions and are different in proton-nucleus as well as proton-proton collisions at LHC energies.
This is interpreted as a hint to sequential meson melting of
heavy vector states, further supported by the suppression of $\Upsilon (3s)$
in heavy-ion collisions \cite{Sirunyan:2018nsz,Aaboud:2018quy,Sirunyan:2018pse,Sirunyan:2017lzi,Aronson:2017ymv}.

Here, we focus on the question whether the DS-BS approach in rainbow-ladder
approximation is capable to deliver Regge type trajectories
of radial excitations when using simple
interaction kernels. To be specific we employ the Alkofer-Watson-Weigel (AWW)
kernel \cite{Alkofer:2002bp} in the pseudo-scalar ($J^{P} =0^{-}$)
and vector ($J^{P} = 1^{-}$) channels
and search for the first excited states. Such a study is a prerequisite for the
extension to non-zero temperatures \cite{Dorkin:2015jck}.
The AWW kernel is known to provide a good description of meson ground states
supposed the analytic properties of the quark propagators are properly
dealt with \cite{Dorkin:2014lxa,Dorkin:2013rsa,Dorkin:2010ut}.
However, in the literature one finds remarks that AWW
is less appropriate for a description of excitations
due to their sensitivity to long-range interactions
\cite{Binosi:2016rxz,Alkofer:2002bp,Mojica:2017tvh,El-Bennich:2016qmb}
(for dedicated studies, cf.\
\cite{Holl:2004fr,Holl:2005vu,Qin:2011xq,Rojas:2014aka,Hilger:2015ora}
for instance).
Nevertheless, we feel that a further investigation is timely, in particular
w.r.t.\ the above stressed importance of Regge trajectories
as an important feature of the meson spectrum.
For the search of meson excitations we employ    a
method, based on investigations of  zeros of  determinants
of the corresponding system of homogeneous equations,
to search for the radial excitations of the BS equation. In our approach,
each value of the mass which zeroes out the determinant above the ground state  is associated with one excited state on the $0^-$ or $1^-$ trajectory.

Our paper is organized as follows. In section 2 we recall the DS and BS
equations as well as the AWW kernel. Numerical results are described
in section 3. We summarize in section 4. The appendix contains
some technicalities.

\section{Recalling the DS and BS equations in rainbow-ladder approximation}
\label{sec:2}
The DS equation (also dubbed gap equation) aims at solving
 \begin{align} \label{eq:DSE}
     S^{-1}(p) \; = \; S^{-1}_0(p) - \int \frac{d^4k}{(2\pi)^4}
\left [-ig^2\gamma^{\nu}\frac{\tau^a}{2}\right ] \mathcal{D}_{\mu\nu}(p,k) \Gamma^{\mu,a}(p,k) S(k),
 \end{align}
for the dressed quark propagator $S$,
where $S_0$ is the   free   quark propagator, $\gamma^{\nu}$ are the Dirac matrices with $\{\gamma^{\mu},\gamma^{\nu}\} = 2\delta^{\mu\nu}$, $\tau^a$ are color matrices, $p$ and $k$ are four-momenta, $g$ is the QCD coupling constant,
and $\mathcal{D}_{\mu\nu}$ stands for the gluon propagator.
In Euclidean space, the rainbow  approximation   consists in a replacement
of the dressed quark-gluon  vertex $\Gamma^{\mu,a}(p,k)$ by the free one,
 $\Gamma^{\mu,a}(p,k) \Rightarrow - i \gamma^{\mu}\frac{\tau^a}{2}$ and in a replacement  of the exact interaction kernel  $g^2\mathcal{D}_{\mu\nu}(k)$ by the free propagator and
   a properly chosen form-factor  $D(k^2)$, i.e,
    $g^2\mathcal{D}_{\mu\nu}(k)\displaystyle\longrightarrow \left(\delta_{\mu\nu} - k_{\mu} k_{\nu} /k^2 \right)D(k^2)$, see below.
The   equation for the quark propagator   reads
 \begin{align} \label{eq:tDSE}
     S^{-1}(p) = S^{-1}_0(p) \; + \; \frac{4}{3} \int\frac{d^4k}{(2\pi)^4} \lbrack g^2 \mathcal{D}_{\mu\nu}(p-k) \rbrack \gamma^{\mu} S(k) \gamma^{\nu}
 \end{align}
with $S^{-1}_0(p) = i \slashed{p} +  m_q$, where the parameter $m_q$
is flavor dependent.

The dressed quark propagator $S(p)$ enters the BS equation for the
vertex function
 \begin{align} \label{eq:BSEgeneral}
     \Gamma(P, p) = \int\frac{d^4k}{(2\pi)^4} K(P, p, k) S(k+\eta_1 P) \Gamma(P,k) S(k-\eta_2 P),
 \end{align}
where  and $p$ are the total and relative momenta of quarks and $\eta_1+\eta_2=1$ describes momentum sharing
$P=(iM_{12},{\bf 0})$ (for a meson   a mass $M_{12}$  at rest), and the rainbow  approximation
for the kernel function
 \begin{align}
    K(P,p,k) = -g^2\mathcal{D}_{\mu\nu}(p-k)\Big( \gamma^{\mu}\frac{\tau^a}{2} \Big) \Big( \gamma^{\nu}\frac{\tau^a}{2} \Big).
 \end{align}
In the Euclidean space, the used BS equation becomes then
 \begin{align} \label{eq:BSEeuclid}
     \Gamma(P,p) = -\frac{4}{3}\int\frac{d^4k}{(2\pi)^4}\gamma^{\mu} S(k +\eta_1 P) \Gamma(P,k) S(k-\eta_2) \gamma^{\nu} \lbrack g^2 \mathcal{D}_{\mu\nu}(p-k) \rbrack.
 \end{align}
  Note the complex valued   momenta  of quarks entering the BS equation (\ref{eq:BSEeuclid}).
In the present paper, we employ   the AWW kernel, i.e.\ $D(k^2) \Rightarrow D^{\mathrm{AWW}}(k^2)$
in the decomposition of the gluon propagator in Landau gauge,
$
g^2\mathcal{D}_{\mu\nu}(k)
= ( \delta_{\mu\nu} - k_{\mu} k_{\nu} k^{-2} ) \, D(k^2)
$
with
 \begin{align} \label{eq:AWWkernel}
     D^{\mathrm{AWW}}(k^2) = \frac{4\pi^2Dk^2}{\omega^2}
e^{-\frac{k^2}{\omega^2}},
 \end{align}
with the interaction strength parameter $D$ and the interaction range parameter $\omega$.
(In what follows, we employ $\eta_1=\eta_2= 1/2$ and the standard
model parameters $\omega = 0.5 \, \mathrm{GeV}$ and $D = 16 \, \mathrm{GeV}^{-2}$,
unless explicitly noted.)
It is the IR part of the Maris-Tandy kernel \cite{Maris:1999nt}.
Results for the AWW and Maris-Tandy kernels are compared
in \cite{Hilger:2017jti}.

\section{Numerical methods and results}\label{sec:3}

\begin{figure}[!tb]
 \centering
\includegraphics[scale=1.2]{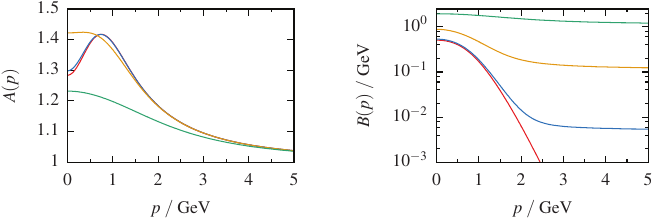} 
\caption{(Color online) Propagator functions
$A(p)$ (left) and $B(p)$ (right) on the posive real axis.
Red curve: $m_q = 0 \, \mathrm{MeV}$ (chiral limit),
blue curve: $m_q = 5 \, \mathrm{MeV}$,
yellow curve: $m_q = 115 \, \mathrm{MeV}$,
green curve: $m_q = 1130 \, \mathrm{MeV}$.}
\label{fig:1}
\end{figure}

The numerical details for solving the above q
uoted DS and BS equations
with given truncations and approximations are described in \cite{Greifenhagen:2016xx}.
In vacuum, the quark propagator can be decomposed as
$S^{-1}(p) = i \gamma \cdot p A(p) + B(p)$ to split (\ref{eq:tDSE}) into two coupled integral equations for
$A$ and $B$ which,  as already mentioned,  are needed for complex arguments $p$.
  The introduced functions $A(p)$ and $B(p)$ are referred to as the quark wave function and
  quark-mass parameter, respectively. The dynamically generated quark mass is then
$M^2_q=\left(B(p) /A(p)\right)^2$.   Figure \ref{fig:1} exhibits examples for these
functions for positive, real values of the momentum $p$. Note the non-linear dependence on the quark mass parameter $m_q$.

Making an expansion of the BS vertex function (\ref{eq:BSEeuclid})
into spin-angular functions, spherical harmonics and Gegenbauer polynomials
  and performing the angular integration explicitly, and
approximating the resulting one-dimensional integrals by a proper quadrature formula,
 one arrives at
   an algebraic system of equations  in a  matrix form
$X_\alpha = \hat S_{\alpha \beta} X_\beta$ with $\alpha, \beta = 1, \cdots N$,
where $N = \alpha_{max} N_{Gegenbauer} N_{Gauss}$. Here, $\alpha_{max}$ denotes the number
of spin-angular harmonics, $N_{Gegenbauer}$ is the number of included Gegenbauer polynomials,
and $N_{Gauss}$ stands for the mesh number of the   employed quadrature formula (Gaussian
 integration mesh, in our case). The
chain of manipulations that leads to the quantity $\hat S_{\alpha \beta}$
is recalled in the Appendix, where also the elements of $X$ are defined,   see also Ref.~\cite{Dorkin:2013rsa} for details.

The energy of mesons as $\bar q q$ bound states is determined by
$\mbox{det} \vert \hat S - \mathbbm{1} \vert = 0$ with $\hat S$ being a function of
the quantity $M_{12}$ which appears in Eq.~(\ref{eq:S}) in the Appendix.
An example is exhibited in Fig.~\ref{fig:2}.

\begin{figure} 
\includegraphics[scale=0.6]{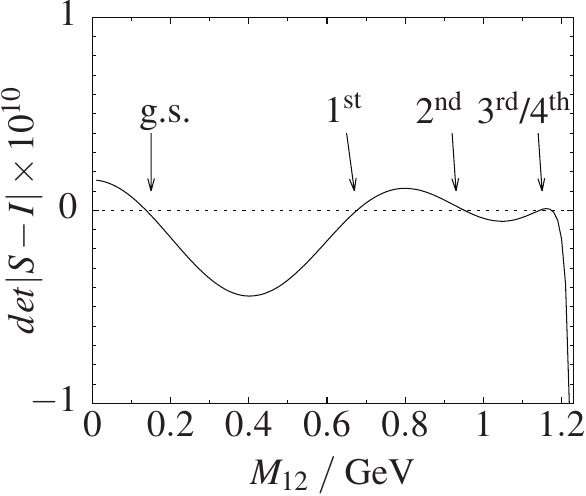}
\caption{Smooth determinant function
$det(\hat S-\mathbbm{1})$ as a function of $M_{12}$ for the pion channel
($m_{1,2} = m_q = m_u = 5 \, \mathrm{MeV}$).
For $\omega = 0.3 \, \mathrm{GeV}$ and $D = 205.761 \, \mathrm{GeV}^{-2}$.
(These non-standard parameters have been chosen for the sake
of a demonstration example which displays higher excitations.)
The arrows denote the masses of ground state (g.s.), first excited state (1\textsuperscript{st}),
second excited state (2\textsuperscript{nd}), third and fourth excited states (3\textsuperscript{rd} and 4\textsuperscript{th}).}
\label{fig:2}
\end{figure}

\begin{figure}[!tb]
 \centering
\includegraphics[scale=0.9]{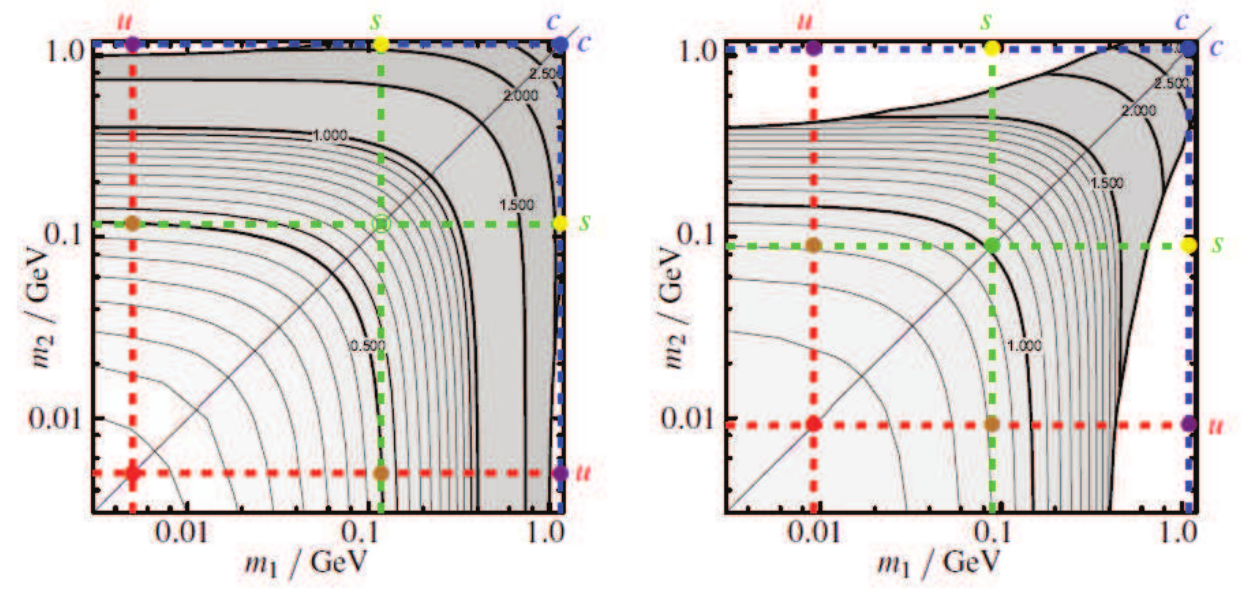}
\caption{ (Color online) Contour plot of pseudo-scalar (left) and vector (right) meson ground
state masses in units of GeV for varying quark masses $m_1$ and $m_2$.
The colored bullets denote the experimental values of meson
ground states (red: $\pi$/$\rho$, green: $K$/$\phi$,
violet: $\eta_c$/$J_{\psi}$) which could be used for extracting
the bare quark mass parameters $m_{1,2}$
(vertical and horizontal dashed lines, labeled by the
corresponding quark flavor).
In the white region, no solutions of the BS equation could found
w/o accounting explicitly for
the pole structure of $S$ in the complex momentum plane.}
\label{fig:3}
\end{figure}

The AWW kernel depends on two parameters, $D$ and $\omega$; in addition the
quark masses $m_{1,2}$ ($m_1 = m_2 = m_q$ for equal quark-mass mesons) must be
adjusted. Figure \ref{fig:3} exhibits examples
of meson ground state masses $M_{n=0, J} (m_1, m_2)$
as contour plots over the $m_1$ -- $m_2$ plane
for the pseudo-scalar (left, $J = 0$) and vector (right, $J = 1$) channels.
One can select suitably three meson masses to determine, for given
$\{ \omega, D \}$, the $m_{u/d, s, c}$ quark masses. For instance,
$M_{\rho, \phi, J/\psi}$ would be such a triple on the $m_1 = m_2$
diagonal, see right panel, or include additionally a non-diagonal
combination, such as $K^*$. A consistency check is provided by a
comparison with quark masses delivered by the $M_{\pi, K, \eta_c}$
mass values by an analog procedure, see left panel.
The off-sets of the dashed horizontal lines in
both panels point to a slight tension, i.e.\ one can not reproduce exactly
the mentioned input meson masses at once for the given interaction
kernel parameters $\{ \omega, D \}$. For instance,
the corresponding value of $m_c =1.110 \, \mathrm{GeV}$
in the vector channel can be compared with one suggested in the pseudo-scalar channel,
with optimum value $1.130 \, \mathrm{GeV}$ etc.

  \begin{table}[ht!] 
   \centering
   \begin{tabular}{lllll|lllll}
    \toprule
\hline
     & our results & results of & \cite{Hilger:2017jti} & \cite{PDG} &  & our results & results of & \cite{Hilger:2017jti} & \cite{PDG}\\
     & $J^{P}=0^{-}$ & $J^{PC}=0^{-+}$ & $J^{PC}=0^{--}$ &  &  & $J^{P}=1^{-}$ & $J^{PC}=1^{--}$ & $J^{PC}=1^{-+}$ & \\
    \midrule
\hline
    $M_{\pi,\mathrm{g.s.}}$ & 0.137 & 0.137 &  & 0.140 & ~$M_{\rho,\mathrm{g.s.}}$ & 0.758 & 0.758 & & 0.775\\
    $M_{\pi,1\mathrm{st}}$ & 0.986 &  & 0.985 &   & ~$M_{\rho,1\mathrm{st}}$ & 1.041 & 1.041 &  & 1.465\\
    $M_{\pi,2\mathrm{nd}}$ & $\textcolor{gray}{1.369^*}$ &  &  & 1.300 & ~$M_{\rho,2\mathrm{nd}}$ & 1.064 &  & 1.062 & \\
     &  &  &  &  & ~$M_{\rho,3\mathrm{rd}}$ & $\textcolor{gray}{1.287^*}$ &  &  & 1.720\\
    \midrule
\hline
    $M_{K,\mathrm{g.s.}}$ & 0.492 & 0.492 &  & 0.494 & ~$M_{K^*,\mathrm{g.s.}}$ & 0.945 & 0.946 &  & 0.894\\
    $M_{K,1\mathrm{st}}$ & 1.162 & 1.162 &  & 1.460  & ~$M_{K^*,1\mathrm{st}}$ & 1.264 & -- &  & 1.414\\
    \midrule
\hline
    $M_{s\overline{s},\mathrm{g.s.}}$ & 0.693 & 0.693 &  &  & ~$M_{\phi,\mathrm{g.s.}}$ & 1.077 & 1.078 &  & 1.019\\
    $M_{s\overline{s},1\mathrm{st}}$ & 1.278 &  & 1.278 &  & ~$M_{\phi,1\mathrm{st}}$ & 1.402 &  & 1.400 & \\
    $M_{s\overline{s},2\mathrm{nd}}$ & $\textcolor{gray}{1.572^*}$ &  &  &  & ~$M_{\phi,2\mathrm{nd}}$ & 1.430 & 1.428 &  & 1.680\\
      & &  &  &  & ~$M_{\phi,3\mathrm{rd}}$ & $\textcolor{gray}{1.598^*}$ &  &  & 2.175\\
    \midrule
\hline
    $M_{D,\mathrm{g.s.}}$ & -- & -- &  & 1.870 & ~$M_{D^*,\mathrm{g.s.}}$ & -- & -- &  & 2.010\\
    \midrule
\hline
    $M_{D_s,\mathrm{g.s.}}$ & 2.075 & 2.041$^\#$ &  & 1.968 & ~$M_{D^*_s,\mathrm{g.s.}}$ & -- & -- &  & 2.112\\
    $M_{D_s,1\mathrm{st}}$ & 2.313 & 2.267$^\#$ &  &  &  &  &  &  & \\
    \midrule
\hline
    $M_{\eta_c,\mathrm{g.s.}}$ & 2.984 & 2.944$^\#$ &  & 2.984 & ~$M_{J/\psi,\mathrm{g.s.}}$ & 3.136 & 3.098$^\#$ &  & 3.097\\
    $M_{\eta_c,1\mathrm{st}}$ & 3.278 &  & 3.225$^\#$ &  & ~$M_{J/\psi,1\mathrm{st}}$ & 3.346 &  & 3.309$^\#$ & \\
    $M_{\eta_c,2\mathrm{nd}}$ & 3.557 & 3.508$^\#$ &  & 3.639 & ~$M_{J/\psi,2\mathrm{nd}}$ & 3.593 & 3.553$^\#$ &  & 3.686\\
      & &  &   &  & ~$M_{J/\psi,3\mathrm{rd}}$ & 3.601 & 3.563$^\#$ &  & 3.773\\
    \bottomrule
\hline
   \end{tabular}
 \caption{Mass spectrum of pseudo-scalar ($J^{P}=0^{-}$) and vector ($J^{P}=1^{-}$) bound states
for the parameter set $\omega = 0.5 \; \mathrm{GeV}$,
$D = 16 \; \mathrm{GeV}^{-2}$,
   $m_u = m_d = 5 \; \mathrm{MeV}$, $m_s = 115 \; \mathrm{MeV}$ and $m_c = 1130 \; \mathrm{MeV}$
and experimental values, in units of GeV. ``g.s.'',``1st'' and ``2nd'' etc.\ stand for
   ground state, the first and second excitations etc. Gray values marked with $^*$ indicate that calculations already reached the pole region.
   The ``--'' for the $D$, $D^*$ and $D^*_s$ ground states
means that no bound state for the employed parameters could be found;
accordingly, there is also no solution for the (radial) excitations.
Note for $D_s$ and $c\overline{c}$ states the different schemes in fixing the quark masses compared to \cite{Hilger:2017jti},
where $m_s = 90 \; \mathrm{MeV}$ and $m_c = 1110 \; \mathrm{MeV}$
are used (marked by ${}^\#$). Experimental data from \cite{PDG}.}
\label{tab:benchmark_ps}
\end{table}

In Table I, we present results of our calculations of the mass spectrum of
ground states and excitations of a few first lightest
pseudo-scalar ($\pi$, $K$, $s\bar s$, $D$ and $\eta$) and vector  ($\rho$, $K^*$,$\phi$, $D_s$ and $J/\Psi$) mesons.
 Whenever possible, our results are compared with experimental
data~\cite{PDG} and with calculations reported in \cite{Hilger:2017jti}.
The lack of corresponding information
 in Table I is denoted by "--". In addition, we note $f_\pi = 0.133 \, \mbox{GeV}$ and
$\langle \bar q q \rangle = (- 0.255 \,\mbox{GeV})^3$~\cite{Greifenhagen:2016xx}.
The overwhelming impression is that, despite of the truncation and
the simple interaction kernel, quite reasonable numbers
for the ground states are delivered
(most notable for $\pi, K, D_s, \eta_c$ in the $0^{-}$ channel
and $\rho, K^*, \phi, J/\psi$ in the $1^{-}$ channel),
however, with some drastic deviations, e.g.\
the pure pseudo-scalar $\bar s s $ states do not  appear in nature
and our failure for $D, D^*, D_s^*$ for the given parameters.
(Employing the parameters of \cite{Hilger:2016efh} we
reproduce accurately the $D, D_s$ results reported there.)
Improvements can be established, e.g.\ to use the full Maris-Tandy
kernel and allow for a flavor dependent variation of the partition
parameter $\eta$ as in Ref.~\cite{Fischer:2014xha,Kubrak:2014ela}. The latter work includes
many more channels from $0^{-+}$ up to $3^{++}$. Focusing on the
natural Regge trajectory sequence $J^{PC} = 1^{--}, 2^{++}, 3^{--}$,~\cite{Fischer:2014xha,Kubrak:2014ela}
find a linear relationship $M_J^2 = M^2(0) + \hat \beta J$ which is intriguing since the approach
does not incorporate any linearly rising inter-quark potential. Due to
our restriction on pseudo-scalar and vector channels we can not make
an analog analysis of orbital Regge trajectories.
Instead, we consider the excitations in the $0^-$ and $1^-$ channels separately
as a generalization of radial Regge trajectories.

\begin{figure}[!htb]
\includegraphics[scale=0.6]{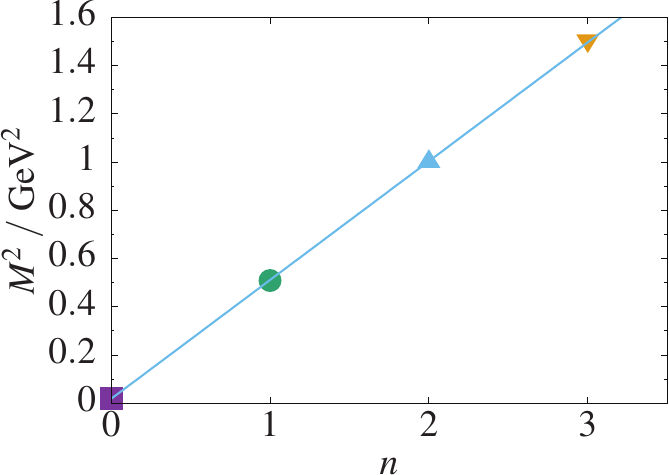}
\put(-81,3){$\widetilde{}$}
\caption{ (Color online) Examples of pion bound sates (symbols, $m_{1,2} = m_q = m_u$)
and the fitted Regge trajectory (blue line) as a function of
the "radial quantum number" $\tilde n$ (which is however a count label for the
zeros of $\mbox{det} \vert \hat S - \mathbbm{1} \vert$)
for $\omega =0.32 \, \mbox{GeV}$ and $a = 0.5 \, \mbox{GeV}^3$.
In contrast to our standard parameters
used in Table I, here the $\tilde n= 1, 2$ states are shifted and an $\tilde n=3$ pion state appears additionally.}
\label{fig:4}
\end{figure}

As already mentioned above,
here, another issue is the discrepancy of calculated and experimental values of
excitations.
Another problem is given by the very few excitations which are accessible
without intruding poles in the relevant complex momentum domain.
In Table I, the states which are hampered by such poles are marked by "*".
Furthermore, determining the bound states by the zeros of
$\mbox{det} \vert \hat S - \mathbbm{1} \vert$ does not strictly ensure
a given $\cal{C}$ parity. Comparing with \cite{Hilger:2017jti}
one observes some states are to be attributed to $J^{PC} = 0^{--}$ and
$J^{PC} = 1^{-+}$, see third columns in Table I in the $0^-$ and $1^-$
parts. Instead of disputing the
issue of exotic \cite{Hilger:2017jti}
or anomalous states
(cf.\ paragraph 3 in \cite{Eichmann:2016nsu} for a comprehensive discussion
as well as \cite{Ahlig:1998qf}),
we take our calculated values of $0^-$ and $1^-$ states
and check the arising sequence of g.s.\ and 1\textsuperscript{st},
2\textsuperscript{nd} $\cdots$ states as proxy for radial
Regge trajectories for linearity, see Fig.~\ref{fig:4} for an example
with apparently $\tilde n$ linear trajectory.
In fact, for certain parameter choices we find such linear Regge trajectories
of excitations parameterized by $M^2 = M_0^2 + \beta \tilde n + c \tilde n^2$
with negligibly small values of $c$.
We ignore thereby that some states have improper $\cal{C}$ parity,
i.e.\ we simply attribute the quantity $\tilde n$ to the count label
of the excited $0^-$ or $1^-$ sates as indicated in Table I.

\begin{figure}[!htb]
 \centering
\includegraphics[scale=0.77]{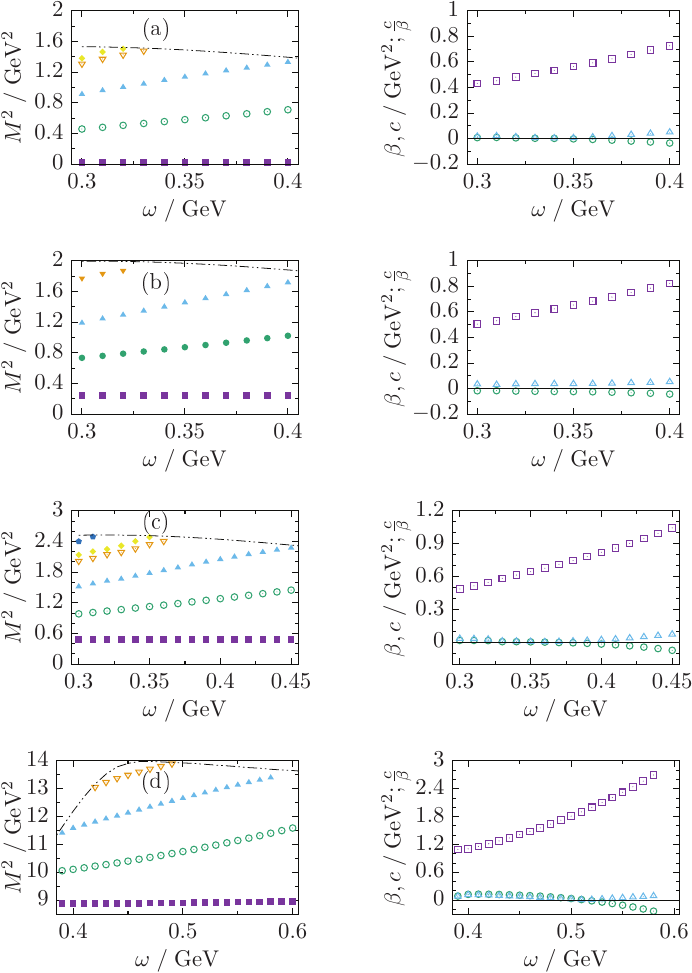}  
\caption{(Color online) Left column:
Spectra of $0^-$ states representing
pions (a: $m_q = m_u$),
kaons (b: $m_1 = m_u$, $m_2= m_s$),
fictitious pseudo-scalar $\bar s s$ states (c: $m_q = m_s$),
and $\eta_c$ (d: $m_q = m_c$)
as a function of $\omega$
for $a \equiv \omega^5 D = 0.5 \, \mbox{GeV}^3$.
The dot-dot-dashed curves mark the limit of the
mass squared region wherein a save determination
without accounting for divergences is possible.
Note the according disappearance of the $\tilde n=3$ excitations
in (a - c) for larger values of $\omega$ at given $a$.
The case of $a = 1$ GeV$^3$ is reported
in \cite{Greifenhagen:2016xx}, where also the (dis)appearance regions
of $D$ and $D_s$ are explored.
Empty symbols: exotic states.
Right column: The corresponding Regge slope coefficients $\beta$
(empty violet squares),
the quadratic term $c$ (empty green circles) in fits of the spectra by
$M_n^2 = M_0^2 + \beta \tilde n +c \tilde n^2$, $\beta \equiv \mu^2$,
and the deviation measure from linear behavior defined by $\vert c / \beta \vert$
(blue triangles).
\label{fig:5} }
\end{figure}

\begin{figure}[!htb]
\includegraphics[scale=0.77]{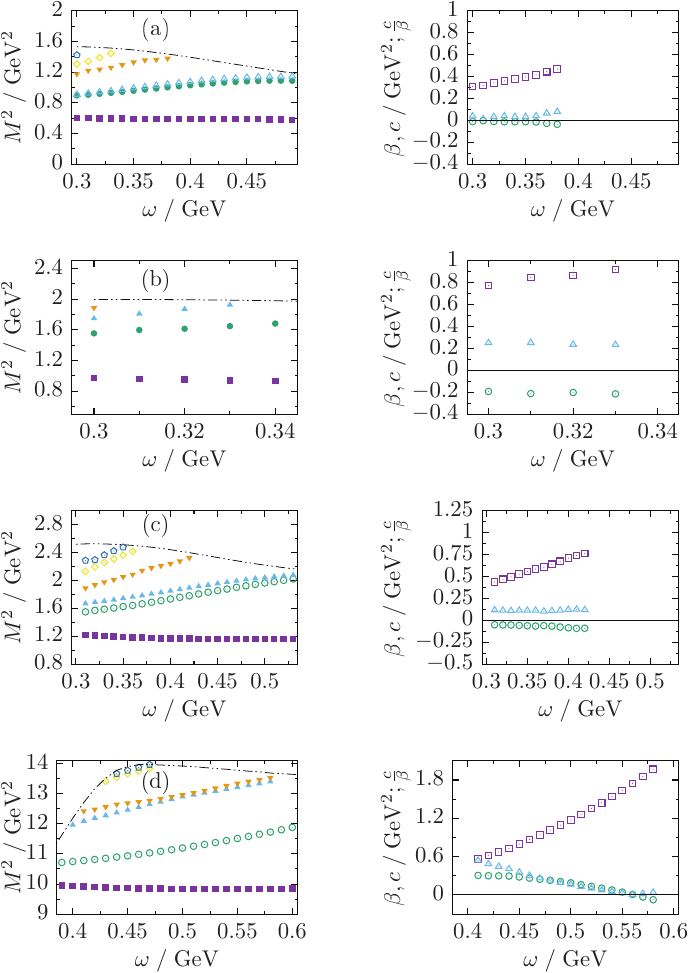}  
\caption{(Color online) As Fig.~\ref{fig:5} but for the $1^-$ vector channel representing
$\rho$ (a: $m_q = m_u$),
$K^*$  (b: $m_1 = m_u$, $m_2= m_s$),
$\phi$ (c: $m_q = m_s$),
and $J/\psi$ (d: $m_q = m_c$).
Note that in the case of very adjacent states these are handled
as one state with averaged energy.
\label{fig:6} }
\end{figure}

Some survey is exhibited in Fig.~\ref{fig:5},
where a few $0^-$ states are depicted (left column) and the Regge slope
parameters as well as a linearity measure are displayed too (right column).
In that study, we freeze in
$a = \omega^5 D $ and vary the parameter $\omega$. As known, the
ground state masses are kept constant under such a variation, but evidently
the excited states depend on $\omega$, even up to a disappearance of certain
states, e.g.\ $\pi$ and $\bar s s $. The slope changes with $\omega$, while
the linearity is strikingly good in the depicted parameter range.
This is in contrast to the $1^{-}$ channel (see Fig.~\ref{fig:6})
where,
at fixed values of $\omega^5 D$, also the ground states vary with changing
$\omega$; linear Regge trajectories are hardly accessible within the preferred
parameter range adjusted to $0^-$ states, cf.\ also~\cite{Mojica:2017tvh}.
One reason is the appearance of very narrow states, similar to the
3rd/4th excitations in Fig.~\ref{fig:2}. The other reason is the large
nonlinearity measure in some cases or both obstacles together,
see Fig.~\ref{fig:6}. Nevertheless, when taking the
averaged energy of the narrow double states and count these
as one state, we do see some Regge type behavior for certain
parameter ranges.

Despite the poor agreement of (radial) excitations with experimental data,
evidenced in Table I
in both the $0^-$ and $1^-$ channels,
we emphasize the occurrence of the linear (Regge) trajectories
w.r.t.\ the count label $\tilde n$ as proxy of the proper radial quantum number
$n$ in the pseudo-scalar channels, thus extending the analysis of
\cite{Fischer:2014xha,Kubrak:2014ela} towards radial excitations.
It remains to be checked whether the Marris-Tandy kernel helps
improving the excitations in the vector channel towards
establishing a linear Regge pattern, at least qualitatively.

\section{Summary}\label{sec:4}

In summary we test the capability to catch the first excited states
of mesons (pseudo-scalar $J^P = 0^-$ and vector $J^P = 1^-$ channels)
by using the
Dyson-Schwinger and Bethe-Salpeter equations in rainbow-ladder
approximation equipped with the Alkofer-Watson-Weigel kernel.
In particular, we establish a certain parameter range in which
excitations do form a linear (Regge) trajectory. This is, however, restricted
to the pseudo-scalar channel. Other channels, including larger angular
momenta, require improvements, among them refined interaction
kernels in conjunction with symmetry preserving truncations.
The ultimate goal is to arrive a coherent framework which catches
the observed sequences of hadron states on Regge trajectories
in both the $J$ and $n$ directions.
This is the first step of an attempt to describe hadron properties,
and thus implicitly confinement and relevant scales, together with
the subsequent extension to finite temperatures and baryon densities
in follow-up investigations.
Irrespectively of the quest for linear Regge trajectories in radial
direction is the need of a proper account of the $1s$, $2s$ and $3s$
states in the quarkonia vector channels $J/\psi$ and $\Upsilon$
which serve as sources of direct probes in ultra-relativistic heavy-ion
collisions.
According
to our contemporary understanding, at some temperature and
at small baryon density, hadrons as quasi-particle degrees of QCD should disappear
in favor of quasi-quark and quasi-gluon degrees of freedom.
The transition happens gradually and may depend on the flavor
channel under consideration. For larger baryon densities, the transition
could be abrupt, supposed a critical point occurs in the phase diagram
of strongly interacting matter. A few large-scale heavy-ion experiments, e.g.\ the
beam-energy scan at RHIC, NA61/SHINE at SPS, CBM at SIS100,
as well at NICA and J-PARC address in their physics programs the
critical point search. For that, both the properties of hadrons as
individual entities and the behavior of hadron matter are key
quantities in reconstructing the final state of strong-interaction matter
in collision.

\begin{acknowledgements}
The authors gratefully acknowledge the collaboration with
S.~M.~Dorkin, T.~ Hilger and M.~Viebach on the topic.
\end{acknowledgements}

\section*{Appendix: Spin-angular harmonics}
\addcontentsline{toc}{section}{Appendix}

  The BS vertex function $\Gamma$ can be expanded into spin-angular harmonics:
  \begin{align} \label{part_amplitudes}
\Gamma(p) = \sum_{\alpha= 1}^{\alpha_{max}}
\Gamma_{\alpha}(p) =  \sum_{\alpha= 1}^{\alpha_{max}}\, g_{\alpha}(p) \, \mathcal{T}_{\alpha}(\vec{p})
\end{align}
with functions $g_{\alpha}$ fulfilling the orthogonality relation
$
     g_{\alpha}(p)  = \int d\Omega_{\vec{p}} \, \Tr\lbrack \, \Gamma(p) \mathcal{T}_{\alpha}^{\dagger}(\vec{p}) \, \rbrack.
$
For pseudo-scalar mesons ($J^{P}=0^{-}$),
the number of independent spin-angular harmonics $\alpha_{max} =4$, and the set is chosen as
  \begin{alignat}{3}
  \begin{aligned} \label{eq:spin_ang_harm_ps}
\mathcal{T}_{1}(\vec{p}) &= \; \frac{1}{\sqrt{16\pi}} \gamma^5
=
\mathcal{T}_{1}^{\dagger}(\vec{p}), \, 
\;\;\;\;\;\;\;\;\;\;\;\; \mathcal{T}_{2}(\vec{p}) &=  \frac{1}{\sqrt{16\pi}} \gamma^0 \gamma^5
= \;
- \mathcal{T}_{2}^{\dagger}(\vec{p}), \\
\mathcal{T}_{3}(\vec{p}) &= \; - \frac{1}{\sqrt{16\pi}} \slashed{n}_{\vec{p}} \gamma^0 \gamma^5
=
\mathcal{T}_{3}^{\dagger}(\vec{p}), \, 
\mathcal{T}_{4}(\vec{p}) &= \; - \frac{1}{\sqrt{16\pi}} \slashed{n}_{\vec{p}} \gamma^5
=
\mathcal{T}_{4}^{\dagger}(\vec{p}),
  \end{aligned}
  \end{alignat}
and for vector mesons ($J^{P}=1^{-}$), $\alpha_{max} = 8$ with
  \begin{alignat}{3}
  \begin{aligned} \label{eq:spin_ang_harm_vec}
     \mathcal{T}_{1}(\vec{p}) &= \; \sqrt{\frac{1}{16\pi}}\slashed{\xi}_{\mathcal{M}}
=
\mathcal{T}_{1}^{\dagger}(\vec{p}), \;
     \mathcal{T}_{2}(\vec{p}) = \; - \sqrt{\frac{1}{16\pi}} \gamma^0 \slashed{\xi}_{\mathcal{M}}
=
\mathcal{T}_{2}^{\dagger}(\vec{p}),\\
     \mathcal{T}_{3}(\vec{p}) &= \; - \sqrt{\frac{3}{16\pi}} (n_{\vec{p}} \, \xi_{\mathcal{M}}) = \mathcal{T}_{3}^{\dagger}(\vec{p}),\\
     \mathcal{T}_{4}(\vec{p}) &= \; \sqrt{\frac{3}{32\pi}} \gamma^0\lbrack -(n_{\vec{p}} \xi_{\mathcal{M}}) + \slashed{n}_{\vec{p}} \, \slashed{\xi}_{\mathcal{M}} \rbrack =-\mathcal{T}_{4}^{\dagger}(\vec{p}),\\
     \mathcal{T}_{5}(\vec{p}) &= \; \sqrt{\frac{1}{32\pi}} \lbrack \slashed{\xi}_{\mathcal{M}} + 3(n_{\vec{p}} \, \xi_{\mathcal{M}})\slashed{n}_{\vec{p}} \rbrack  =- \mathcal{T}_{5}^{\dagger}(\vec{p}), \\
     \mathcal{T}_{6}(\vec{p}) &= \; \sqrt{\frac{1}{32\pi}} \gamma^0 \lbrack \slashed{\xi}_{\mathcal{M}} + 3(n_{\vec{p}} \, \xi_{\mathcal{M}})\slashed{n}_{\vec{p}} \rbrack = \mathcal{T}_{6}^{\dagger}(\vec{p}), \\
     \mathcal{T}_{7}(\vec{p}) &= \; - \sqrt{\frac{3}{16\pi}} \gamma^0 (n_{\vec{p}} \, \xi_{\mathcal{M}})
= \mathcal{T}_{7}^{\dagger}(\vec{p}), \\
     \mathcal{T}_{8}(\vec{p}) &= \; \sqrt{\frac{3}{32\pi}} \lbrack -(n_{\vec{p}} \, \xi_{\mathcal{M}}) + \slashed{n}_{\vec{p}} \, \slashed{\xi}_{\mathcal{M}} \rbrack =- \mathcal{T}_{8}^{\dagger}(\vec{p}).
   \end{aligned}
   \end{alignat}
Scalar products are displayed here in Minkowski space; $n_{\vec{p}}$ is the unit vector $n_{\vec{p}} = (0, \vec{p}/|\vec{p}|)$,
$\xi_{\mathcal{M}} = (0,\vec{\xi_{\mathcal{M}}})$  is the polarization vector fixed by $\vec{\xi}_{+1} = -(1,i,0)/\sqrt{2}$, $\vec{\xi}_{-1} = (1,-i,0)/\sqrt{2}$,
$\vec{\xi_{0}} = (0,0,1)$ and slashed quantities such as $\slashed{x}$
represent $\gamma^{\mu}x_{\mu}$.

The partial amplitudes $\Gamma_{\alpha}(p)$ and the interaction kernel \eqref{eq:AWWkernel}
are decomposed over the basis of spherical harmonics $Y_{lm}(\theta,\phi)$ and normalized Gegenbauer polynomials $X_{nl}(\chi)$,
  \begin{align}
     Z_{nlm} &= X_{nl}(\chi) Y_{lm}(\theta,\phi) \\
             &= \sqrt{\frac{2^{2l+1}}{\pi}\frac{(n+1)(n-l)!l!^2}{(n+l+1)!}} \sin^l\chi G^{l+1}_{n-l}(\cos\chi) Y_{lm}(\theta,\phi),
  \end{align}
with familiar Gegenbauer polynomials $G^{l+1}_{n-l}(\cos\chi)$. The hyper angle $\chi$ is defined by $\cos \chi = p_4/\tilde{p}$ and $\sin \chi = |\vec{p}|/\tilde{p}$,
where $\tilde{p} = (p_4^2 + \vec{p}^{\,2})^{1/2}$ is the modulus
for an Euclidean four-vector $p = (p_4,\vec{p})$.
The partial decompositions of $\Gamma_{\alpha}(p)$ and $D^{AWW}(p-k)$
read
  \begin{align} \label{eq:part_decomp_gamma}
     \Gamma_{\alpha}(p) &= \sum_{n} \varphi^n_{\alpha,l_{\alpha}}(\tilde{p}) X_{nl_{\alpha}}(\chi_p) \mathcal{T}_{\alpha}(\vec{p}) , \\
     \label{eq:part_decomp_D} D^{AWW}(p-k) &= 2\pi^2 \sum_{\kappa\lambda\mu} \frac{1}{\kappa + 1} V_{\kappa}(\tilde{p},\tilde{k}) X_{\kappa\lambda}(\chi_p) X_{\kappa\lambda}(\chi_k) Y_{\lambda\mu}(\Omega_p) Y^*_{\lambda\mu}(\Omega_k),
  \end{align}
  where $V_{\kappa}(\tilde{p},\tilde{k})$ are the partial kernels and $\varphi^n_{\alpha,l_{\alpha}}(\tilde{p})$ are the expansion coefficients of the partial amplitudes.
  Actually $l_{\alpha}$ is restricted by the corresponding orbital momentum encoded in $\mathcal{T}_{\alpha}(\vec{p})$. For $\mathcal{T}_{1,2}(\vec{p})$ from
Eq.~\eqref{eq:spin_ang_harm_ps},
  $l_{\alpha} = 0$ holds,
while for $\mathcal{T}_{3,4}(\vec{p})$ $l_{\alpha} = 1$. In analogy for vector mesons (see \eqref{eq:spin_ang_harm_vec}), $l_{\alpha} = 0$ for $\mathcal{T}_{1,2}(\vec{p})$,
  $l_{\alpha} = 1$ for $\mathcal{T}_{3,4,7,8}(\vec{p})$ and $l_{\alpha} = 2$ for $\mathcal{T}_{5,6}(\vec{p})$.

The spin-angular harmonics  $\mathcal{T}_{\alpha}(\vec{p})$
possess a well defined ${\cal C}$ parity, and the normalized Gegenbauer polynomials
$X_{nl}(\chi_p)$ have a well defined symmetry too. Therefore, keeping only each
second entry in the sum in Eq.~(\ref{eq:part_decomp_gamma}) yields a well defined
${\cal C}$ parity of the bound states upon
$\bar \Gamma (q, P) \equiv [ {\cal C} \, \Gamma (-q, P) \, {\cal C}^{-1}]^T
= \eta_C \Gamma (q, P)$ with ${\cal C} = \gamma_2 \gamma_4$
(see \cite{Hilger:2017jti} for further details).

Changing the integration variables to the hyperspace,
$
d^4k = \tilde{k}^3\sin^2\chi_k \sin\theta_k d\tilde{k} 
$
$\times d\chi_k d\theta_k d\phi_k$,
  inserting Eq. \eqref{eq:part_decomp_gamma} and \eqref{eq:part_decomp_D} into \eqref{eq:BSEeuclid} and performing the necessary angular integration, a system of integral
  equations for the expansion coefficients $\varphi^n_{\alpha,l_{\alpha}}(\tilde{p})$
as the $N$ elements of $X$ remains:
  \begin{align}  \label{eq:phi}
     \varphi^n_{\alpha,l_{\alpha}}(\tilde{p}) = \sum_{\beta} \sum_{m=1}^{\infty} \int d\tilde{k} \tilde{k}^3 \hat S_{\alpha \beta}(\tilde{p},\tilde{k},m,n) \varphi^m_{\beta,l_{\beta}}(\tilde{k}).
  \end{align}
  The explicit expression for $\hat S_{\alpha \beta}(\tilde{p},\tilde{k},m,n)$ reads
  \begin{align}
  \begin{aligned} \label{eq:S}
   \hat S_{\alpha \beta}(\tilde{p},\tilde{k},m,n)
&= \sum_{\kappa} \int \sin^2\chi_k d\chi_k X_{ml_{\beta}}(\chi_k) X_{\kappa \lambda}(\chi_k) \sigma_{s,v}(\tilde{k}_1^2)\sigma_{s,v}(\tilde{k}_2^2)\\
& \times A_{\alpha \beta}(\tilde{p},\tilde{k},\kappa,\chi_k,n),
  \end{aligned}
  \end{align}
where $\tilde{k}_{1,2}^2$ is given by
$\tilde k_{1,2}^2 = \tilde k^2 - \eta_{1,2}^2 M_{12}^2
\pm \eta_{1,2} M_{12} \cos \chi$
with momentum partitioning parameters $\eta_1 + \eta_2 = 1$
and
$A_{\alpha \beta}(\tilde{p},\tilde{k},\kappa,\chi_k,n)$
results from calculations of traces and
angular integrations as
\begin{align}
  \begin{aligned} \label{eq:A}
    A_{\alpha \beta}(\tilde{p},\tilde{k},\kappa,\chi_k,n)
&= \int \sin^2\chi_p d\chi_p d\Omega_p d\Omega_k V_{\kappa}(\tilde{p},\tilde{k})
X_{nl_{\alpha}}(\chi_p) X_{\kappa \lambda}(\chi_p) Y_{\lambda\mu}(\Omega_p) Y^*_{\lambda\mu}(\Omega_k) \\
& \times \Tr[d_{\mu \nu}((p-k)^2) \gamma^{\mu}...\mathcal{T}_{\alpha}(\vec{p})
\cdots \mathcal{T}_{\alpha}(\vec{p}) \gamma^{\nu}].
  \end{aligned}
\end{align}


\begin{thebibliography}{99.}%

\bibitem{Yang:2016sws}
  Z.~Yang, Q.~Wang and U.~G.~Meissner,
  Phys.\ Lett.\ B {\bf 767} (2017) 470.

\bibitem{Anisovich:2000kxa}
  A.~V.~Anisovich, V.~V.~Anisovich and A.~V.~Sarantsev,
  Phys.\ Rev.\ D {\bf 62} (2000) 051502.

\bibitem{Masjuan:2012gc}
  P.~Masjuan, E.~Ruiz Arriola and W.~Broniowski,
  Phys.\ Rev.\ D {\bf 85} (2012) 094006.

\bibitem{Masjuan:2013xta}
  P.~Masjuan, E.~Ruiz Arriola and W.~Broniowski,
  Phys.\ Rev.\ D {\bf 87} (2013) 118502.

\bibitem{Klempt:2007cp}
  E.~Klempt and A.~Zaitsev,
  Phys.\ Rept.\  {\bf 454} (2007) 1.

\bibitem{Brodsky:2014yha}
  S.~J.~Brodsky, G.~F.~de Teramond, H.~G.~Dosch and J.~Erlich,
  Phys.\ Rept.\  {\bf 584} (2015) 1.

\bibitem{Bugg:2012yt}
  D.~V.~Bugg,
  Phys.\ Rev.\ D {\bf 87} (2013) 118501.

\bibitem{Bugg:2004xu}
  D.~V.~Bugg,
  Phys.\ Rept.\  {\bf 397} (2004) 257.

\bibitem{Afonin:2006wt}
  S.~S.~Afonin,
  Eur.\ Phys.\ J.\ A {\bf 29} (2006) 327.

\bibitem{Afonin:2016wie}
  S.~S.~Afonin and I.~V.~Pusenkov,
  EPJ Web Conf.\  {\bf 125} (2016) 04006.

\bibitem{Ebert:2009ua}
  D.~Ebert, R.~N.~Faustov and V.~O.~Galkin,
  Eur.\ Phys.\ J.\ C {\bf 66} (2010) 197.

\bibitem{Ebert:2009ub}
  D.~Ebert, R.~N.~Faustov and V.~O.~Galkin,
  Phys.\ Rev.\ D {\bf 79} (2009) 114029.

\bibitem{Afonin:2009xi}
  S.~S.~Afonin,
  Phys.\ Lett.\ B {\bf 675} (2009) 54.

\bibitem{Afonin:2009pd}
  S.~S.~Afonin,
  Phys.\ Lett.\ B {\bf 678} (2009) 477.

\bibitem{Li:2013oda}
  D.~Li and M.~Huang,
  JHEP {\bf 1311} (2013) 088.

\bibitem{Li:2012ay}
  D.~Li, M.~Huang and Q.~S.~Yan,
  Eur.\ Phys.\ J.\ C {\bf 73} (2013) 2615.

\bibitem{Zollner:2017ggh}
  R.~Z\"ollner and B.~K\"ampfer,
  J.\ Phys.\ Conf.\ Ser.\  {\bf 1024} (2018)  012003.

\bibitem{Fischer:2014xha}
  C.~S.~Fischer, S.~Kubrak and R.~Williams,
  Eur.\ Phys.\ J.\ A {\bf 50} (2014) 126.

\bibitem{Hilger:2017jti}
  T.~Hilger, M.~Gomez-Rocha, A.~Krassnigg and W.~Lucha,
  Eur.\ Phys.\ J.\ A {\bf 53} (2017)  213.

\bibitem{Binosi:2016rxz}
  D.~Binosi, L.~Chang, J.~Papavassiliou, S.~X.~Qin and C.~D.~Roberts,
  Phys.\ Rev.\ D {\bf 93} (2016) 096010.

\bibitem{Wang:2013wk}
  K.~l.~Wang, Y.~x.~Liu, L.~Chang, C.~D.~Roberts and S.~M.~Schmidt,
  Phys.\ Rev.\ D {\bf 87} (2013)  074038.

\bibitem{Suganuma:2017syi}
  H.~Suganuma, T.~M.~Doi, K.~Redlich and C.~Sasaki,
  J.\ Phys.\ G {\bf 44} (2017) 124001.

\bibitem{Sirunyan:2018nsz}
  A.~M.~Sirunyan {\it et al.} [CMS Collaboration],
  Phys.\ Lett.\ B {\bf 790}, 270 (2019).

\bibitem{Aaboud:2018quy}
  M.~Aaboud {\it et al.} [ATLAS Collaboration],
  Eur.\ Phys.\ J.\ C {\bf 78}, no. 9, 762 (2018)

\bibitem{Sirunyan:2018pse}
  A.~M.~Sirunyan {\it et al.} [CMS Collaboration],
  Phys.\ Lett.\ B {\bf 790}, 509 (2019).

\bibitem{Sirunyan:2017lzi}
  A.~M.~Sirunyan {\it et al.} [CMS Collaboration],
  Phys.\ Rev.\ Lett.\  {\bf 120}, no. 14, 142301 (2018).

\bibitem{Aronson:2017ymv}
  S.~Aronson, E.~Borras, B.~Odegard, R.~Sharma and I.~Vitev,
  Phys.\ Lett.\ B {\bf 778}, 384 (2018).

\bibitem{Alkofer:2002bp}
  R.~Alkofer, P.~Watson and H.~Weigel,
  Phys.\ Rev.\ D {\bf 65} (2002) 094026.

\bibitem{Dorkin:2015jck}
  S.~M.~Dorkin, M.~Viebach, L.~P.~Kaptari and B.~K\"ampfer,
  J.\ Mod.\ Phys.\  {\bf 7} (2016) 2071.

\bibitem{Dorkin:2014lxa}
  S.~M.~Dorkin, L.~P.~Kaptari and B.~K\"ampfer,
  Phys.\ Rev.\ C {\bf 91} (2015)  055201.

\bibitem{Dorkin:2013rsa}
  S.~M.~Dorkin, L.~P.~Kaptari, T.~Hilger and B.~K\"ampfer,
  Phys.\ Rev.\ C {\bf 89} (2014) 034005.

\bibitem{Dorkin:2010ut}
  S.~M.~Dorkin, T.~Hilger, L.~P.~Kaptari and B.~K\"ampfer,
  Few Body Syst.\  {\bf 49} (2011) 247.

\bibitem{Mojica:2017tvh}
  F.~F.~Mojica, C.~E.~Vera, E.~Rojas and B.~El-Bennich,
  Phys.\ Rev.\ D {\bf 96} (2017)  014012.

\bibitem{El-Bennich:2016qmb}
  B.~El-Bennich, G.~Krein, E.~Rojas and F.~E.~Serna,
  Few Body Syst.\  {\bf 57} (2016)  955.

\bibitem{Holl:2004fr}
  A.~Holl, A.~Krassnigg and C.~D.~Roberts,
  Phys.\ Rev.\ C {\bf 70} (2004) 042203.

\bibitem{Holl:2005vu}
  A.~Holl, A.~Krassnigg, P.~Maris, C.~D.~Roberts and S.~V.~Wright,
  Phys.\ Rev.\ C {\bf 71} (2005) 065204.

\bibitem{Qin:2011xq}
  S.~x.~Qin, L.~Chang, Y.~x.~Liu, C.~D.~Roberts and D.~J.~Wilson,
  Phys.\ Rev.\ C {\bf 85} (2012) 035202.

\bibitem{Rojas:2014aka}
  E.~Rojas, B.~El-Bennich and J.~P.~B.~C.~de Melo,
  Phys.\ Rev.\ D {\bf 90} (2014) 074025.

\bibitem{Hilger:2015ora}
  T.~Hilger, M.~Gomez-Rocha and A.~Krassnigg,
  Eur.\ Phys.\ J.\ C {\bf 77} (2017)  625.
%

%
\bibitem{Maris:1999nt}
  P.~Maris and P.~C.~Tandy,
  Phys.\ Rev.\ C {\bf 60} (1999) 055214.

\bibitem{Greifenhagen:2016xx}
R.~Greifenhagen,
Investigation of the AWW kernel for describing the excited meson spectrum
in a combined Dyson-Schwinger -- Bethe-Salpeter approach,
Master Thesis, TU Dresden (2016).

\bibitem{PDG}
M. Tanabashi et al. (Particle Data Group),
Phys. Rev. D {\bf 98} (2018)  030001.

\bibitem{Hilger:2016efh}
  T.~Hilger and A.~Krassnigg,
  Eur.\ Phys.\ J.\ A {\bf 53} (2017) 142.

\bibitem{Kubrak:2014ela}
  S.~Kubrak, C.~S.~Fischer and R.~Williams,
  J.\ Phys.\ Conf.\ Ser.\  {\bf 599} (2015) 012013.

\bibitem{Eichmann:2016nsu}
  G.~Eichmann,
  Few Body Syst.\  {\bf 58}, no. 2, 81 (2017)

\bibitem{Ahlig:1998qf}
  S.~Ahlig and R.~Alkofer,
  Annals Phys.\  {\bf 275}, 113 (1999)

\end{thebibliography}
\end{document}